# Summary of a Literature Review in Scalability of QoS-aware Service Composition

Leticia Duboc
Dept. of Media Technologies La Salle - University Ramon Lull
Spain
lduboc@salleurl.edu

Faisal AlRebeish
King Abdulaziz City for Science and Technology
Riyadh, Saudi Arabia
frebeish@kacst.edu.sa

Vivek Nallur
Distributed Systems Group Trinity College
Dublin, Ireland
vivek.nallur@scss.tcd.ie

Rami Bahsoon
Dept. of Computer Science
University of Birmingham
United Kingdom
r.bahsoon@cs.bham.ac.uk

*Abstract* — Dynamic Service composition is a process by which a service-based application can compose itself, based on multiple requirements like functional specifications, QoS requirements, and cost constraints. Mechanisms to discover and evaluate potential services, while optimizing QoS, is a NP-hard problem. Hence, most solutions focus on obtaining a good selection of services that meets the QoS constraints of an application. One would expect that, in such a scenario, the problem of scalability of dynamic service composition would be well understood. Nevertheless, this paper shows that authors have no consistent way to characterize the scalability of their solutions, and so consider only a limited number of scaling characteristics. This review aimed at establishing the evidence that the route for designing and evaluating the scalability of dynamic QoS-aware service composition mechanisms has been lacking systematic guidance, and has been informed by a very limited set of criteria. For such, we analyzed 47 papers, from 2004 to 2018.

## I. STUDY DESCRIPTION

This template, We followed the guidelines of Kitchenham[1] for systematic literature reviews in Software Engineering.

### A. RESEARCH QUESTIONS

Our study intended to answer the following questions:

[RQ1:] How are authors on QoS-based dynamic service composition evaluating the scalability of their solutions?

[RQ2:] What are the scaling dimensions (and their value ranges) being considered by researchers in this area to characterize the scalability of their solutions?

[RQ3:] What are the metrics being considered by researchers to characterize the scalability of their solutions?

### B. DATA SOURCES AND SEARCH STRATEGY

In software engineering, electronic databases are normally considered sufficient [2][3]. Hence, we used the following search engines: IEEE Explore, ACM Digital Library, Science Direct, Engineering Village (which searches INSPEC as well as EI Compendex), ISI Web of Knowledge, Google Scholar, and CiteSeer. We searched all the data sources, using a combination of the following keywords (and variations): quality of service (QoS, QoS-aware, QoS-enabled), web service composition (WSC, Service Composition, Service-based, Service-Oriented, Service-based Architecture, Service oriented Architecture, Service-selection) and dynamic (Adaptive, Adaptation, Self-adaptive, Self-optimizing, Self-healing, Self-managing). The search strings formed by these keywords were adapted for each search engine.

### C. STUDY SELECTION

Papers were selected for the analysis based on the following inclusion and exclusion criteria: [Inclusion] Conference papers, journal articles, workshop papers and technical reports. [Exclusion] Criteria E1-E7 helped to filter out miscategorized results, or results that were only tangentially related to the topic . Criteria E8-E11, applied after full text screening, aim at selecting papers whose techniques can be compared against each other. Criteria is presented with a summarised rationale when judged necessary.

[E1:] Papers unrelated to dynamic service selection; [E2:] Papers that were completely domain-specific. Rationale: we seek to identify attributes and metrics relevant to the problem of dynamic composition in general; [E3:] Papers with journal extensions (latter included); [E4:] Papers prior the Web Service standard (year 2000); [E5:] Papers not published in English; [E6:] Duplicate references; [E7:] Papers unobtainable from databases or authors; [E8:] Papers dealing exclusively with technical improvements to the underlying infrastructure (e.g., SOAP, WSDL, BPEL, DAML-S, OWL-S), with no mention of QoS. Rationale: these technologies do not address the problem of matching services based on QoS.

[E9:] Papers without a mechanism for QoS evaluation. Rationale: necessary for selecting services for the workflow. Using cost as the only determiner is insufficient; [E10:] Papers with no mention of workflow or its abstract services. Rationale: service-based applications are composed in a workflow, which gives rises to its end to end QoS. Mechanisms that do not consider a workflow or its tasks structure are outside our scope; [E11:] Papers with no mention of candidate services for an abstract service. Rationale.

Our problem involves choosing the right service from many candidates. Papers that did not address this were not within scope. In order to apply criteria E8-E11, we unified the vocabulary used by the papers, defining (i) an abstract services the functional specification of a certain task (sometimes referred to as a task or service class or abstract service in the application workflow); and (ii) a candidate service as an implementation of an abstract service (also referred to as service candidate or concrete service). Each candidate service has a QoS that it advertises through its SLA. Following the application of all exclusion criteria, 47 papers were selected for analysis (listed in Appendix 1).

*D. Data Extraction*

The following data was extracted from each article:

• Title, authors, publication venue and year;

• Scalability claimed for the solution?

• Claim based on evaluation?

• Approach used for evaluation of scalability.

• Metrics considered for characterizing scalability.

• QoS characteristic considered for service composition and their constraints (when specified).

• Scaling dimensions considered for evaluation of scalability and their range of values (when specified).

• Technique for composition of services.

• Optimization/utility function for composing services.

*E. Threats to Validity*

The study had the following threats to validity:

*Construct validity*: Our study aimed at understanding how the scalability of QoS-based dynamic service composition was being evaluated. One concern is related to the appropriateness of measures to answer our research questions. We believe the measures collected are sufficient to provide a fair characterization of the state of-the-art of scalability analysis according to Duboc et al.'s [4].

*Internal validity*: Main threats are incomplete and/or wrong selection of primary studies and individual researchers' bias. These threats were mitigated by following a pre-defined protocol, carrying out several dry runs individually, and consolidating the differences collaboratively. The selection of primary studies and data extraction was performed individually by three researchers, with another researcher serving as a third-party control. Values inferred from graphs in papers are approximate, due to their low resolution. Hence, while values are internally consistent, they are not necessarily exact. External validity: Our scope covered only academic data sources, automatically precludin any commercial solution, not indexed by these sources. Also, our exclusion criteria excludes domain-specific papers and papers relating to improvements in SOAP, WSDL, OWL, etc.

## II. SUMMARY OF THE RESULTS

After Answers to our research questions follows. We emphasize that we do not evaluate the scalability of each solution. Rather, we survey the research landscape for techniques that have been used to claim their scalability.

*[RQ1.] How are authors on QoS-based dynamic service composition evaluating the scalability of their solutions?*

Scalability is the ability of a system to maintain the satisfaction of its quality goals to levels that are acceptable to its stakeholders when characteristics of the application domain and the system design vary over expected operational ranges [4]. Therefore, any analysis of an ordinary software quality in the presence of the variation of characteristics of the application domain and system design is, in fact, a scalability analysis [5]. We have observed that, out of 47 papers, 41 present some form of scalability analysis or claims. Some are explicit, as in paper [P2] which has a section entitled "QoSMOS Scalability, where authors analyse execution time with respect to the number of concrete services and abstract medical services. A similar section is presented in paper[P39],where authors evaluate the execution time of the algorithm with respect to the number of concrete services, abstract services and QoS constraints for a service composition. Other works present varied analyses without an explicit mention of the term "scalability". This is the case of the paper [P10], which test for the percentage of optimal solutions found given an increasing number of services to combine.

The remaining 6 papers ([P22], [P25] ,[P26] ,[P28], [P38], [P40]) provide only working examples of their solutions, without accounting for variation. For example, paper [P24] shows the utility values for 8 abstract service and 40 candidate services. Only one paper [P28] does not give any numeric example at all. Our review also shows that papers which presented claim of scalability or of software quality (without consideration for variation), based their claims on some sort of evaluation: 33 papers use simulations, while the remaining 14 have used testing or a working example.

*[RQ2.] What are the scaling dimensions (and their value ranges) being considered by researchers in the area of QoS based dynamic service composition to characterize the scalability of their solutions?*

Our review showed that papers considered a large variety of scaling dimensions, pertaining both to the application domain and system design. Nearly all papers considered the workflow size and the number of candidate services (39 and 41 papers respectively) in their analysis. Other 19 application domain scaling dimensions have been mentioned, such as the number of QoS attributes, the number of requests per day, the amount of data transmission between services, the number of breach of services, among others. With the exception of the number of QoS, which were mentioned by three papers, all others were listed by at most one paper.

It is interesting to note that authors can be very specific with respect to the scaling dimensions. Paper [P14], for example, differentiates between the number of QoS attributes and the number of user QoS constraints, meaning that though there may be many QoS attributes that are measured, there might be only a few that the user is concerned about. In these cases, the search space can be shrunk dramatically.

Value ranges considered in the evaluations vary even more greatly. Take for example, the workflow size, which varied from 5 to 10000, considering all papers. Nevertheless, each paper evaluated its own range. For example, paper [P2] considered the range [5 - 16], while paper [P15] tested for [10 - 100]. Only one work evaluated workflows with more than 100 services, this was paper [P4].

Regarding the number of candidate services per abstract service, the range considered across the papers varied from 1 to 10000. As with the workflow size, different papers considered different ranges. With respect to the system design, 16 dimensions have been mentioned; some using a numeric scale.

Paper [P1], for example, varied the number of ants in a ant colony optimization (ACO)algorithm along the values 3, 6 and 7. Paper [P3] varied the max non-improving generations (MNIG), a design variable determining the termination condition, from 1000 to 5000. Others represented thresholds, such as the threshold for a statistical model of volatility in [P16]. Finally, some dimensions represented design choices, such as the choice of building tree algorithm in [P12] and the configuration mode in [P6], which varied among global dynamic, local dynamic and static.

*[RQ3.] What are the metrics being considered by researchers in the area of QoS-based dynamic service composition to characterize the scalability of their solutions?*

Papers also varied the metrics used to evaluate the scalability of their solution. Most papers (30 out of 47) evaluated the impact of the scaling dimensions against execution time, such as [P1], [P15], [P26] and [P43]. Four papers used success and failure rates as their metric of choice ([P6][P8][P11][P26]). Papers [P9],[P14], [P19] and P[39]. looked into the utility achieved against the optimal result, while papers [P12] and [P17] considered costs.

Other metrics, mentioned by at most one paper, were fitness value of the algorithm, speedup, average of violated quality of service, time complexity, percentage of optimal solutions, interruption time caused by reselection process, overall aggregated QoS, memory usage, availability. More than one metric was used to evaluate the solution in 25 out of 47 papers. [Other observations] Works vary with respect to the technique for dynamic services composition. In the papers, 27 different techniques and 7 optimization/utility functions were used. From the optimization functions, simple additive weighting was the most popular (29 papers). The QoS considered for the composition also varied: 23 different qualities were mentioned. Most solutions adopted execution time (40 papers), followed by cost (29 papers), availability (28 papers) and reliability (27 papers). Some others were: throughput, popularity/reputation, success rate, composability, , maintainability, eco-impact and quality of the documentation. The number of Qos also varied. Most papers, 40 out of 47, used between 2 and 6 QoS. Only two, [P9] and [P7], used a greater number of QoS, 9 and 10 respectively. Six papers did not specify the QoS considered.

## III. DISCUSSION OF THE RESULTS

The results above confirm that scalability is indeed a concern for Qos-based service dynamic composition. Most authors attempt to justify their (explicit or not) scalability claims with evaluations of software qualities given some variation in the application domain and system design. Nevertheless, these evaluations vary greatly. With respect to the scaling dimensions, there is a general agreement that the workflow size and the number of candidate services should be considered. Some works also agree on the number of QoS as an important scaling dimension. However, for all other application domain quantities, each paper has its particular concern. As for t he scaling dimensions belonging to the system design, there is no consensus on variables. This is only natural, as each study has its own approach for service composition. The variables ranges of values also vary widely. With respect to commonly used variables, such as the workflow size and the number of candidate services, the variation in ranges were of orders of magnitude. Regarding the metrics used to measure the scalability of solutions, most agreed that execution time was an important concern, followed by success/failure rates, utility and costs. The study also revealed many other, less popular, metrics. Furthermore, each work assessed their own combination of metrics.

All these differences in the scalability are by no means surprising, as in the literature review we could also observe the variety of techniques used for service composition. The fact that there is no single set of metrics and scaling dimensions that fit the scalability analysis of all QoS-based dynamic composition papers is consistent with our view of scalability [4].

The problem with this observation, however, is that authors working on dynamic service composition might be at a loss when planning a scalability evaluation of their solutions, or when wishing to compare the scalability of their approach with others in the literature. What these authors are missing is a systematic mechanism for the selection of metrics and scaling dimensions for their scalability analyses. Such a mechanism must be able to identify with respect to a particular solution, aspects of the application domain and system design that may affect its scalability, and which software qualities might be affected by these characteristics. Using ad-hoc approaches entails the risk of overlooking relevant variables for the analysis.

# APPENDIX A
## PAPERS SELECTED FOR THE LITERATURE REVIEW 1-2

| | | |
|---|---|---|
| [P1] Wang X.; Jing Z.; Yang H. 2011. Service Selection Constraint Model and Optimization Algorithm for Web Service Composition. Information Technology Journal | [P2] Calinescu R.; Grunske L.; Kwiatkowska M.; Mirandola R.; Tamburrelli T.; 2011. Dynamic QoS Management and Optimization in Service-Based Systems. IEEE Trans. Softw. Eng. | [P3] Li S.; Chen M. 2010. An adaptive-GA based QoS driven service selection for Web services composition, In Intl. Conf. on Computer Application and System Modeling (ICCASM) |
| [P4] Ardagna D.; Mirandola R. 2010. Per-flow optimal service selection for Web services based processes. J. Syst. Softw. | [P5] Liu Z.; Xiao R. 2010. Towards Service Selective Optimizing Based on Key Business Process Performance. In 2nd Intl. Conf. on e-Business and Information System Security (EBISS) | [P6] Yang H.; Li Z. 2010. Improving QoS of Web Service Composition by Dynamic Configuration. Information Technology Journal |
| [P7] Li J.; Zhao Y.; Liu M.; Sun Hd; MaD. 2010. An adaptive heuristic approach for distributed QoS-based service composition,In IEEE Simp. on Computers and Communications (ISCC) | [P8] Sheu R.; Lo W.; Lin C.; and Yuan S. 2010. Design and Implementation of a Relaxable Web Service Composition System. In Intl. Conf. on on Cyber-Enabled Distributed Computing and Knowledge Discovery (CYBERC '10) | [P9] Alrifai M.; Risse T. 2009. Combining global optimization with local selection for efficient QoS-aware service composition. In 18th Intl. Conf. on World wide web (WWW '09) |
| [P10] Chen Z.; Wang H.; Pan P. 2009. An Approach to Optimal Web Service Composition Based on QoS and User Preferences. In Intl. Joint Conf. on Artificial Intelligence (JCAI '09) | [P11] Guoping Z.; Huijuan Z.; Zhibin W. 2009. A QoS-Based Web Services Selection Method for Dynamic Web Service Composition. In First Intl. Workshop on Education Technology and Computer Science (ETCS '09) | [P12] Hristoskova, A.; Volckaert, B.; De Turck, F. 2009. Dynamic Composition of Semantically Annotated Web Services through QoS-Aware HTN Planning Algorithms, In 4th Intl. Conf. on Internet and Web Applications and Services. |
| [P13] Yang, L.; Dai, Y.; Zang B.. 2009. Performance Prediction Based EX-QoS Driven Approach for Adaptive Service Composition, J. Inf. Sci. Eng. | [P14] Chen Z; Yao Q. 2008. A Framework for QoS-aware Web Service Composition in Pervasive Computing Environments,In 3rd Intl. Conf. on Pervasive Computing and Applications. | [P15] Ko, J. M.; Kim, C. O. & Kwon, I.-H. 2008. Quality-of-service oriented web service composition algorithm and planning architecture. In Journal of Systems and Software |
| [P16] Chafle, G.; Doshi, P.; Harney, J.; Mittal, S.; Srivastava, B. 2007. Improved Adaptation of Web Service Compositions Using Value of Changed Information, In IEEE Intl. Conf. on Web Services (ICWS'07) | [P17] Eckert, J.; Repp, N.; 0002, S. S.; Berbner, R. & Steinmetz, R. 2007. An Approach for Capacity Planning of Web Service Workflows. in John A. Hoxmeier & Stephen Hayne, ed., 'AMCIS', Association for Information Systems | [P18] Wancheng N.; Lingjuan H.; Lianchen L.; Cheng W. 2007. Commodity-Market Based Services Selection in Dynamic Web Service Composition. In 2nd IEEE Asia-Pacific Service Computing Conference |
| [P19] Mohabey, M.; Narahari, Y.; Mallick, S.; Suresh, P.; Subrahmanya, S. V. 2007. A Combinatorial Procurement Auction for QoS-Aware Web Services Composition. In IEEE Intl. Conf. on Automation Science and Engineering. | [P20] Yang Y.; Tang S; Xu Y.; Zhang W.; Fang L. 2007. An Approach to QoS-aware Service Selection in Dynamic Web Service Composition. In 3rd Intl. Conf. on Networking and Services (ICNS '07). | [P21] Yu T.; Zhang Y.; Lin K. 2007. Efficient algorithms for Web services selection with end-to-end QoS constraints. ACM Trans. |
| [P22] Berbner, R.; Spahn, M.; Repp, N.; Heckmann, O. & Steinmetz, R. 2006, An Approach for Replanning of Web Service Workflows., in Guillermo Rodrguez-Abitia & Ignacio Ania B., ed., 'AMCIS', Assoc. for Information Systems | [P23] Gao A.; Yang D.; Tang S.; Zhang M. 2005. Web Service Composition Using Integer Programming-based Models. In IEEE Intl. Conf. on e-Business Engineering (ICEBE '05) | [P24] Yu T.; Lin K. 2005. A broker-based framework for QoS-aware Web service composition. In IEEE Intl. Conf. on e-Technology, e-Commerce and e-Service (EEE '05) |
| [P25] Canfora G.; Penta M.; Esposito R.; Villani M. 2005. QoS-Aware Replanning of Composite Web Services. In IEEE Intl. Conf. on Web Services (ICWS '05) | [P26] Zeng L.; Benatallah B.; Ngu A., Dumas M.; Kalagnanam J.;Chang H. 2004. QoS-Aware Middleware for Web Services Composition. IEEE Trans. Softw. Eng. | [P27] Menasce D. A. 2004. Composing Web Services: A QoS View. In IEEE Internet Computing |
| [P28] Aggarwal R.; Verma K.; Miller J;, Milnor W. 2004. Constraint Driven Web Service Composition in METEOR-S. In IEEE Intl. Conf. on Services Computing (SCC '04) | [P29] Badidi E.; Esmahi L.; Serhani M. A. 2005. A Queuing Model for Service Selection of Multi-classes QoS-aware Web Services. In 3rd European Conf. on Web Services (ECOWS '05) | [P30] Boone B.; Hoecke S. V.; Seghbroeck G. V.; Joncheere N.; Jonckers V.; Turck F. D.; Develder C.; Dhoedt B. 2010. SALSA: QoS-aware load balancing for autonomous service brokering. J. Syst. Softw. |
| [P31] Guo H.; Huai J.; Li H.; Deng T.; Li Y.; Du Z.. 2007. ANGEL: Optimal Configuration for High Available Service Composition. In IEEE Intl. Conf. on Web Services, (ICWS'07) | [P32] Serhani M. A.; Dssouli R.; Hafid A.; Sahraoui H. 2005. A QoS Broker Based Architecture for Efficient Web Services Selection. In IEEE Intl. Conf. on Web Services (ICWS '05) | P[33]Q. Wu and Q. Zhu, Transactional and QoS-aware dynamic service composition based on ant colony optimization, 2013. Future Generation Computer Systems, vol. 29, no. 5. |
| P[34]F. Mardukhi, N. NematBakhsha, K. Zamanifar and A. Barati, QoS decomposition for service composition using genetic algorithm,. 203. Applied Soft Computing, vol. 13. | P[35]B. Huang, C. Li and F. Tao, A chaos control optimal algorithm for QoSbased,. 2014. Enterprise Information Systems, vol. 8, no. 4, pp. 445-463. | P[36]S. Wang, Q. Sun, H. Zou and F. Yang, Particle Swarm Optimization with Skyline Operator for Fast Cloud-based Web Service Composition,. 2013. Mobile Networks & Appl., vol. 18, no. 1. |

## PAPERS SELECTED FOR THE LITERATURE REVIEW 2-2

| | | |
|---|---|---|
| P[37] W. Dou, X. Zhang, J. Liu and a. J. Chen, HireSome-II: Towards Privacy-Aware Cross-Cloud Service Composition for Big Data Applications. , 2015. Parallel and Distributed Systems, IEEE Transactions on, vol. 26, no. 2. | P[38] A. Klein, F. Ishikawa and S. Honiden, SanGA: A Self-Adaptive Network-Aware Approach to Service Composition. 2014 Services Computing, IEEE Tran, vol. 7, no. 3, pp. 452-464. | P[39] I. Paik, W. Chen and M. N. Huhns, A Scalable Architecture for Automatic Service Composition. 2014. Services Computing, IEEE Tran, vol. 7, no. 1, pp. 82-95. |
| P[40] A. A. P. Kazema, H. Pedramb and H. Abolhassanic, BNQM: A Bayesian Network based QoS Model for Grid service composition. 2015. Expert Systems with Applications, vol. 42, no. 20, p. 68286843. | P[41] H. Jin, X. Yao and Y. Chen, Correlation-aware QoS modeling and manufacturing cloud service composition, Journal of Intelligent Manufacturing. 2015. pp. 1-14. | P[42] J. Lartigaua, X. Xua, L. Niea and D. Zhana, Cloud manufacturing service composition based on QoS with geo-perspective transportation using an improved Artificial Bee Colony optimisation algorithm, 2015. International Journal of Production Research, vol. 53, no. 14, pp. 4380-4404. |
| P[43] C. Jatoth, G.R. Gangadharan, U. Fiore and R. Buyya, QoS-aware Big service composition using MapReduce based evolutionary algorithm with guided mutation 2017. Future Generation Computer Systems | P[44] S. Deng, H. Wu, D. Hu and J. Leon Zhao, Service Selection for Composition with QoS Correlations IEEE Transactions on Services Computing. 2016. vol. 9, no. 2, pp. 291-303. | P[45] X. Peng, J. Gu, t. Tan, Tian ,J. Sun,Y. Yu, B. Nuseibeh, and W. Zhao, CrowdService: Optimizing Mobile Crowdsourcing and Service Composition 2018. ACM Trans. Internet Technol, vol. 18, no. 2, pp. 1-25. |
| P[46] J. Chengfeng , L. Miao and K. Xiang, Edge cloud computing service composition based on modified bird swarm optimization in the internet of things 2018. Cluster Computing. | P[47] H. Ying , Q. Peng , Z. Jiyou , F. Dajuan and P. Huanfeng, Multi-objective service composition model based on cost-effective optimization Applied Intelligence. 2017. vol. 48, no. 3, pp. 651669. | |